\documentclass[aps, twocolumn, prb, amsmath, superscriptaddress, 8pt]{revtex4-2}

\usepackage{graphicx}
\usepackage{gensymb}
\usepackage{color}
\usepackage[dvipsnames]{xcolor}
\usepackage{float}
\usepackage{upgreek}
\usepackage{bm}
\usepackage{amsmath,amssymb}
\usepackage[colorlinks=true,citecolor=blue,linkcolor=blue]{hyperref}

\begin{document}

\title{Reconstruction of non-trivial magnetization textures from magnetic field images using neural networks}

\author{David A. Broadway}
\email{david.broadway@rmit.edu.au}
\affiliation{School of Science, RMIT University, Melbourne, VIC 3001, Australia}
\affiliation{Department of Physics, University of Basel,  Basel, Switzerland}

\author{Mykhailo Flaks}
\affiliation{Department of Physics, University of Basel,  Basel, Switzerland}

\author{Adrien E. E. Dubois}
\affiliation{Department of Physics, University of Basel,  Basel, Switzerland}
\affiliation{QNAMI AG, Hofackerstrasse 40 B, Muttenz CH-4132, Switzerland}

\author{Patrick Maletinsky}
\affiliation{Department of Physics, University of Basel,  Basel, Switzerland}

\begin{abstract} 
Spatial imaging of magnetic stray fields from magnetic materials is a useful tool for identifying the underlying magnetic configurations of the material. 
However, transforming the magnetic image into a magnetization image is an ill-poised problem, which can result in artefacts that limit the inferences that can be made on the material under investigation. 
In this work, we develop a neural network fitting approach that approximates this transformation, reducing these artefacts. 
Additionally, we demonstrate that this approach allows the inclusion of additional models and bounds that are not possible with traditional reconstruction methods. 
These advantages allow for the reconstruction of non-trivial magnetization textures with varying magnetization directions in thin-film magnets, which was not possible previously. 
We demonstrate this new capability by performing magnetization reconstructions on a variety of topological spin textures. 

\end{abstract}

\maketitle

\section{Introduction}
Material science requires efficient and quantitative methods that can ascertain the underlying properties of materials. 
Bulk material measurement techniques have greatly improved our understanding of material properties on a macro scale. 
However, when these materials have a reduced dimensionality and are in the ultra-thin limit (2D and near 2D), where the material is approaching a single atomic layer, these properties can change dramatically. 
For various materials that exhibit this behavior (such as graphene) such novel properties can be probed through electrical measurements, however, these detect the average response of the device, thus missing crucial information about spatial variation. 
In contrast, another class of 2D materials is magnetic and has relevant properties that cannot necessarily be probed through electrical means. 
Despite this, 2D magnetic materials are promising candidates for novel technology and are of great scientific interest~\cite{yuan_quantum_2022, ahn_progress_2024, wang_magnetic_2022}.
As such, it is important to have reliable techniques~\cite{christensen_2024} to determine the underlying magnetization properties of these materials on the micro to nanoscale\,\cite{Thiel2019, song_direct_2021, huang_revealing_2023, li_observation_2024, sun_magnetic_2021, ghiasi_nitrogen-vacancy_2023, rizzo_visualizing_2022, zur_magnetic_2023, tschudin_imaging_2024, robertson_imaging_2022, huang_layer-dependent_2023, scholten_multi-species_2023, li_proximity-induced_2023, grover_chern_2022}.

Imaging the stray magnetic fields from these materials is extremely useful for understanding the magnetization strength, direction, and texture. 
However, inferring this information from the magnetic images alone can be difficult and prone to misinterpretation. 
Alternatively, it is possible to reconstruct the magnetization image by calculating the transformation matrix (from Maxwell's Equations with the assumption that the magnetization is confined to a quasi-2D layer) in Fourier space~\cite{Tan1996, lima2009, Casola2018}, such that
\begin{equation}
    \mathcal{B} = \mathcal{A}\mathcal{M},
\end{equation}
where $\mathcal{B}$ and $\mathcal{M}$ are the Fourier space representation of the magnetic field and magnetization respectively, and $\mathcal{A}$ is a linear map between the two. 
However, in general, the matrix $\mathcal{A}$ is non-invertible, thus, the inverse transformation from magnetic field to magnetization is ill-posed and requires an approximation of the matrix $\mathcal{A^\prime} \approx \mathcal{A}^{-1}$.
This approximation can be made for both
current density and out-of-plane magnetization, where the ill-poised component is trivial (and confined to the Fourier vector $k =  0$, therefore corresponding to the DC field).
As such, reconstructing current density~\cite{Roth1989, Knauss2001, Tetienne2019, Broadway2020a, Ku2020, jenkins_imaging_2022, aharon-steinberg_direct_2022, Chang2017, palm_observation_2024, palm_imaging_2022, chen_current_2024} and out-of-plane magnetization~\cite{Broadway2020a, Broadway2020, Thiel2019, Dovzhenko2018, grover_chern_2022, Appel2019, sun_magnetic_2021, song_direct_2021} has been broadly adopted.
In the case of in-plane magnetization, the number of undefined terms increases, leading to an increase in artifacts and amplification of noise~\cite{Broadway2020a}, and thus, such reconstruction has not been broadly performed thus far.

An alternative approach to approximating the inverse transformation is to use neural networks (NN), which have been shown to perform well for other ill-posed inverse problems~\cite{raissi_physics-informed_2019, Lucas2018a, Prato2008, Ongie2020, Khoo2019, Arridge2019}. 
These techniques require large datasets for training the NN which is not always applicable for materials studies, and are intrinsically biased by the dataset used for training. 
To overcome these restrictions, Dubois et al.~\cite{dubois_untrained_2022} developed an approach in which the physically informed NN learns on a single image by comparing the magnetic field calculated from its output magnetization to the experimental data to generate a Loss function, akin to a fit. 
This approach attempts to approximate the inverse transformation $\mathcal{A}^{-1}$ such that the magnetization image output is a valid (i.e. consistent with Maxwell's equations) source for the measured magnetic field. 
In that work, it was demonstrated that the NN method can outperform the traditional Fourier transformation method.  
Due to these advantages, NN approaches to reconstruction have been used on multiple occasions both for magnetization~\cite{li_puzzling_2023, tschudin_imaging_2024} and current density~\cite{chen_current_2024, reed_machine_2024} reconstruction. 
Additionally, Bayesian based optimization approaches have also been used demonstrating similar advantages~\cite{Clement2019, midha_optimized_2024, yao_universal_2024}. 
However, these previous NN works were still limited to having a uniform magnetization direction within a single image -- a constraint comparable to the one of the Fourier method.

In this work, we demonstrate that NN reconstruction can be used to determine the magnetization from samples with varying magnetization directions within a single image; an important step toward the reconstruction of complex magnetic textures. 
We demonstrate that physically informed limits can be placed on these NNs leading to further improvement in magnetic reconstructions compared with traditional approaches. 
Additionally, we show further improvement by using the full magnetic field vector in these reconstructions, which leads to a more robust magnetization estimation.
Finally, we demonstrate this improved NN reconstruction method on a series of magnetic Skyrmions of differing characters and demonstrate the ability to differentiate between Bloch and N\'eel type Skyrmions and can even identify the helicity of Bloch Skyrmions.

\section{Neural network reconstruction method}

\begin{figure*}
    \centering
    \includegraphics{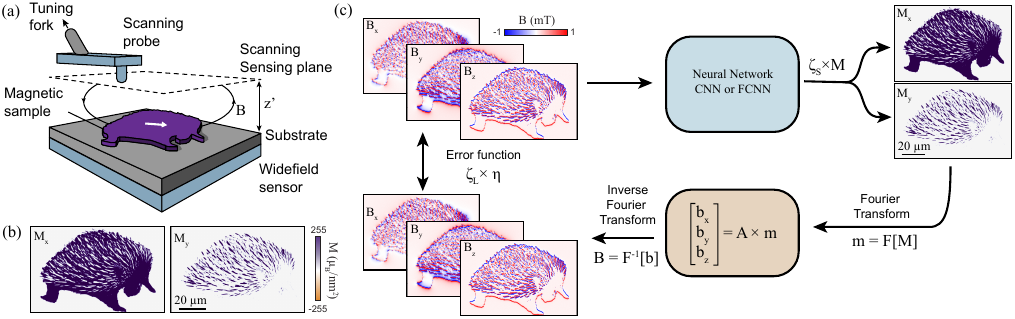}
    \caption{Neural network reconstruction method.
    (a) Schematic of a magnetic sample being imaged by either a scanning probe magnetometer or a widefield magnetometer.
    (b) Simulated in-plane magnetization in the shape of an Echidna with the body having a $M_x$ magnetization and the quills having a $M_y$ magnetization. 
    (c) Schematic of the neural network reconstruction method with the simulated magnetic fields from (b). }
    \label{fig:arch}
\end{figure*}

Reconstruction of a source quantity from a magnetic field is a common practice for many different types of magnetic sensors, including scanning probes and wide-field imagers. 
While there are differences in the spatial resolution, typical stand-off distances, and sensitivity between scanning probes and widefield magnetic imaging, the challenges in the magnetization reconstruction process and the approach to NN-based reconstruction are similar. 
In all cases, a magnetic image is taken in a sensing plane offset from the magnetic sample by a distance $z^\prime$ (Fig.~\ref{fig:arch}a), is fed into the NN, that produces an estimate of the source magnetization, that is then forward-propagated into a magnetic field through the application of the well-poised transformation in Fourier space, and is finally compared to the original magnetic field input as the error function. 
To illustrate the NN reconstruction process we have simulated a magnetic sample in the shape of an Echidna, where the body has a $M_x$ magnetization and the quills have a $M_y$ magnetization (Fig.~\ref{fig:arch}b). 
This magnetization is then forward-propagated into a single magnetic field or $B_{x,y,z}$ field components to be fed into a NN for reconstruction (Fig.~\ref{fig:arch}c).
Depending on the type of sensor, the magnetic field component measured will vary. 
However, it is common to transform this magnetic image into Fourier space and to decompose the image into the three vector components $B_x$, $B_y$, and $B_z$~\cite{Casola2018, Tetienne2019}, allowing the same reconstruction procedure independent of the magnetic sensor type or orientation.
This versatile approach has distinct advantages that we will discuss in Sec.\,\ref{sec:multi}.

\section{Enforcing physically informed bounds}\label{sec:bounds}

A key issue both the traditional Fourier and NN reconstructions face is the allocation of nonzero magnetization to background regions of the sample that contain no magnetic material (e.g. to regions outside of finite-sized "flakes" of 2D magnetic van der Waals materials). 
In some cases, the attribution of background magnetization is approximately homogenous, allowing the background to be subtracted.
However, when other magnetic material is nearby (potentially just outside the field of view) this can lead to an incorrect estimation of the sample's magnetization within the image, as the stray fields can overlap.
This is worse when the sample is only partially captured within the image, as the accompanying, significant loss of information introduces artifacts.
In the case of out-of-plane magnetization and current density, reflective boundary conditions significantly help minimise boundary truncation artifacts\,\cite{Meltzer2017}. 
However, this boundary condition has an additional complication with in-plane magnetization as the inversion of adjacent maps in the Meltzer boundary method\,\cite{Meltzer2017} depends on the sensor and magnetization direction.
Consequently, it is important to introduce more explicit, physically informed bounds on the magnetization itself, which is not possible with the Fourier method but is with the NNs. 

A trivial bound for magnetization is the absolute value lying predominantly along a given axis, that is, $M^\prime_{NN} = \left| M_{NN} \right|$, where $M_{NN}$ is the raw output of the NN and $M^\prime_{NN}$ is the bounded output.
This bound is very effective at minimising attribution of magnetization to the non-sample background, as this often exhibits the inverse sign to the magnetization within the sample. 
However, this approach is only practical for a subset of magnetic images, where the magnetization direction does not vary greatly (less than 90$^\circ$).
In the case of more complex magnetic patterns, the singular sign bound is no longer valid.
Thus, it is important to have generally applicable bounds that can improve the magnetization reconstruction universally. 


\begin{figure}
    \centering
    \includegraphics{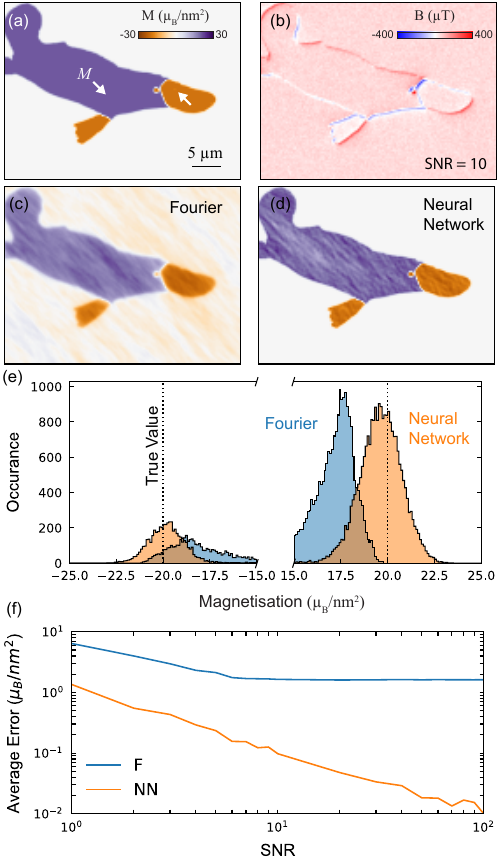}
    \caption{Improved reconstruction with masks. 
    (a) Simulation of magnetization in the shape of a platypus with a magnetization direction of $(\theta, \phi) = (80^\circ, -45^\circ)$ with a partial boundary crossing of the shape. 
    (b) Calculated magnetic field along a sensor direction $(\theta, \phi) = (54.7^\circ, -45^\circ)$ from the simulation in (a) with SNR = 10.
    (c,d) magnetization reconstruction from (b) using the Fourier method (c) and the NN method (d). 
    (e) Histogram of the reconstructions where the Fourier reconstruction has had the mean background subtracted. The true simulated values are shown as dashed lines.
    (f) Comparison of the average error within the platypus of the two methods compared to the true value.
    F stands for the Fourier method and NN is for the Neural Network reconstruction.}
    \label{fig:mask}
\end{figure}

The NN approach allows for the inclusion of weighted masks at various stages that can improve the reconstruction process, leading to a more accurate magnetization reconstruction by minimising artifacts and reducing the attribution of magnetization to erroneous regions.  
In particular, one can include a weighting to the magnetization such that,
\begin{equation}
    M^\prime_{NN} = \zeta_S  M_{NN}
\end{equation}
where $M_{NN}$ is the output of the NN and $M^\prime_{NN}$ is the magnetization after applying the weighted mask, and $\zeta_S$ is a weighted mask, whose values range from 0 to 1 and effectively constrains the region where magnetization will be reconstructed. 
The magnetization mask is applied after each training epoch before being transformed back to the magnetic field components and can be hard (i.e., contains only 0 or 1) or soft (varying values).
This weighting style can be used to penalise magnetization reconstruction in specific regions without prohibiting it. 
However, the reconstruction is more reliable if the magnetization is limited completely when the sample geometry and magnetization region is \textit{a priori} known (as is the case, e.g. in patterned magnetic films, or "flakes" of 2D vdW magnets). 

There are a few practical ways in which one can reliably define a mask. 
One is to define the mask by finding the edges of the material with a threshold or edge-finding algorithm, which can programmatically define a mask for a given image. 
Alternatively, one can draw the mask by tracing the material shape with a series of connected points.
In either case, both approaches or others will commonly have the issue that the mask is not perfectly defined (i.e., is pixel perfect).
To accommodate for this imperfection, the hard mask can be convolved with a Gaussian or other line profile to produce soft edges.
This ultimately results in a blurring of the sharp edges, which can introduce a decrease in the quantitative estimation of the magnetization but can be a useful trade-off when required.

To demonstrate the utility of these masks, we compare the reconstruction quality of the NN approach with a mask to the Fourier-based method. 
We have simulated an in-plane magnetic sample with some canting giving a magnetization direction of $(\theta, \phi) = (80^\circ, -45^\circ)$, where $\theta$ is the longitudinal angle and $\phi$ is the azimuthal (Fig.\,\ref{fig:mask}a). 
While in scanning magnetic systems at larger fields, NV magnetometry (a common magnetic imaging technology\,\cite{Rondin2012,Thiel2019, scholten_widefield_2021}) requires the field to be aligned with the diamond crystal axis, which commonly results in the magnetic field having an angle $\theta \approx 55^\circ$ from the sample normal. 
The out-of-plane component of such bias magnetic fields can induce some canting of the in-plane magnetization and thus is a reasonable scenario to compare~\cite{healey_varied_2022, dubois_untrained_2022}. 
While the NN is agnostic to the field direction, the Fourier method has noise amplification issues for pure in-plane magnetization\,\cite{Broadway2020a}, as such, the out-of-plane component helps to reduce transformation artifacts in the Fourier method. 
By including this canting the Fourier method result is improved while the NN is left unchanged.

To compare with real measurements, we add to the calculated magnetic field a noise background (Fig.\,\ref{fig:mask}b -- there, with a signal-to noise-ratio (SNR) of $10$). 
A comparison of the raw output of the Fourier method (with reflective boundaries) and the NN with a mask, is shown in Fig.\,\ref{fig:mask}c and d respectively. 
While the Fourier method will attribute some background magnetization, in the absence of magnetic signal from other magnetic material contributing from outside the field of view, the mean of this background can be subtracted to get a more accurate estimation of the magnetic moments in the material. 
However, while the background magnetization can be normalised, the magnetization is still underestimated in the Fourier method, while the neural network approach is far more accurate also in this regard (Fig.\,\ref{fig:mask}e).
To illustrate this difference, we calculated the average difference in the reconstructed magnetization and the ground truth as a function of SNR (Fig.\,\ref{fig:mask}f), which shows that the NN approach always outperforms the Fourier method. 
We note that the Fourier method can return a reliable estimate of the magnetisation for out-of-plane magnetized samples that are contained completely within the field of view with no spurious magnetic sources.

In some cases, a source mask may be impossible or  difficult to define, for example because the imaged are is completely filled with magnetised material or because edges are fuzzy or ill-defined.  
In these circumstances, it can be beneficial to apply a different type of weighting that will still focus the NN optimization algorithm on the desired region but doesn't completely restrict the allocation of magnetization in space. 
This alternative is a weighted mask applied to the loss function itself such that,
\begin{equation}
    \eta^\prime = \zeta_L \eta =  \zeta_L\sum_i \left| B_i - B_i^\prime \right|
\end{equation}
where the error $\eta$ is the sum of absolute differences between the measures map $B$ and the reconstructed map $B^\prime$, and $\zeta_L$ is the weighting of the loss function. 
This mask type has the benefit of being less strictly defined as it does not directly restrict the magnetization but rather the region in which the error is calculated. 
However, it can also allow for the inclusion of nonphysical magnetization distributions in these regions and as such, a combination of both masks is useful. 
For example, a loss mask can be useful for focusing the NN reconstruction on specific regions of the image with higher SNR, which can be useful for wide-field imaging setups where the beam shape of the laser illumination defines such regions with higher SNR.
Additionally, in the case of significantly truncated data, a common approach to improve the Fourier transformations is to pad or extend the datasets with simulations (used for both reconstruction and forward transformation as it improves the transformation to Fourier space itself which both approaches use).
In these situations, limiting the error calculation to the measured data region can help improve the convergence of the NN optimization algorithm, as the simulated/extended region often has no noise which can lead to the NN focusing on fitting the extended data and under-fitting the measured data.

\section{Improved reconstruction via multi-component fitting}\label{sec:multi}

\begin{figure}
    \centering
    \includegraphics{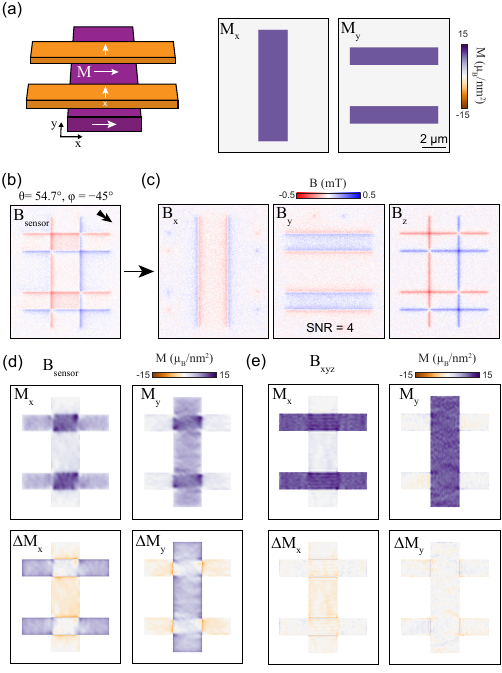}
    \caption{Multiple magnetic field component fitting. 
    (a) Simulation of two overlapping magnetic samples whose in-plane magnetic direction is along the $x$ and $y$ axes. 
    (b) Calculated magnetic field images with an SNR of 4 for a singular sensor direction $(\theta = 54.7^\circ, \phi = -45^\circ$) 
    (c) Decomposition of the single magnetic field direction image into $B_{xyz}$.
    (d) Reconstruction of $M_{x}$ and $M_{y}$ using the $B_{sensor}$ image (top panels) and the difference between the reconstruction and the true value (bottom panels).
    (e) Same as (d) but using all three decomposed magnetic field images.
    }
    \label{fig:multi b}
\end{figure}

Previous work~\cite{dubois_untrained_2022} showed that NN reconstruction can outperform standard Fourier networks by minimising noise amplification and by including the magnetization direction as a fit parameter.
That work, however, was restricted to a single magnetization direction. 
Here, we introduce a second output channel of the neural network (shown in Fig.~\ref{fig:arch}), which allows for the simultaneous determination of both $M_x$ and $M_y$ magnetization components. 
Applying the NN reconstruction for a varying magnetization direction within an image comes at a computational and complexity cost (e.g. the solution phase space has increased the number of deep local minima) leading to a higher likelihood of the NN returning an erroneous magnetization image.
As such, it is important to improve the reliability of the NN for such cases where the magnetization direction is allowed to be inhomogeneous.

Beyond masks, the NN optimization algorithm can be further improved by including additional information as an input to the network.  
A single magnetic field component image ($B_{\rm sensor}$) can readily be transformed into all three vector component ($B_{x},B_{y},$ and $B_{z}$) images via Fourier decomposition~\cite{lima2009, Casola2018, Tetienne2019}.  
While in principle the maps of $B_{x}, B_{y},$ and $B_{z}$ contain no additional information compared to the $B_{\rm sensor}$ map, including them can stabilize the optimization algorithm of the neural network. 
To illustrate this, we consider the case of an in-plane magnetized material with uniform magnetization. 
If one were to consider just the $B_z$ component, then there exist three trivial magnetization distributions (ignoring more complex magnetization textures) that can lead to the same stray field signature, 
\begin{enumerate}
    \item Uniform magnetization within the sample.
    \item Uniform background magnetization of a magnetization direction opposite to the previous case.
    \item Varying magnetization confined to the edges of the sample. 
\end{enumerate}
The first two scenarios are just the inverted allocation of the magnetization which produces identical stray edge fields at the barrier between magnetic material and the background.
The last scenario can also replicate this edge field distribution by varying the magnetization at the edges to replicate the field distribution, often involving sub-resolution variation in the magnetization direction.
In principle, there is no unique solution to the field distribution so the same field can be produced by more complex magnetization patterns.
These solutions can often be ruled out as nonphysical given knowledge of the material. 
In practice, the NN can converge to any of these solutions as they are equally valid, but an unbound NN will converge to a mixture of all three. 
Including a source mask will eliminate the magnetization outside of the sample but this still leaves the issue of edge magnetization. 
Considering all three magnetic field components, $B_{x}$, $B_{y}$, and $B_z$, helps minimise the inclusion of incorrect magnetization distributions (e.g. edge magnetization), as we will show in the following.
This multi-component method approach is illustrated in Fig.\,\ref{fig:arch}(c), where the NN is modified to take three input channels rather than one, and the forward magnetic field calculation is adjusted to match.

To demonstrate the difference between reconstruction with a single component and all three components, we simulate two overlapping magnetic samples that have a magnetization direction that is orthogonal and is aligned to the xy-plane in the shape of a hallbar (Fig.~\ref{fig:multi b}a). 
The magnetic field with noise (here, with SNR = 4) for a single magnetic sensor direction is simulated (Fig.~\ref{fig:multi b}b, $\theta = 54.7$, $\phi = -45$) at a standoff distance of $100~$nm (all typical for scanning NV magnetometry, and decomposed into the $x$, $y$, and $z$-components (Fig.~\ref{fig:multi b}c). 
Next, we reconstructed $M_{x}$ and $M_{y}$ using both the simulated $B_{\rm sensor}$ image (Fig.~\ref{fig:multi b}d) and the multi-component method (Fig.~\ref{fig:multi b}e), for a total of $500$ epochs.

The differences between the single and multi component approaches are highlighted in the difference between the simulated magnetization and reconstruction (Fig.~\ref{fig:multi b}e,d, bottom panels). 
While the reconstruction based on $B_{\rm sensor}$ alone captures the overall features of the sample, there is a correlation between the images where magnetization has been incorrectly assigned to $M_{x}$ instead of $M_{y}$ and vice versa, as well as having more noise reflected in the reconstruction. 
In contrast, the multi-component reconstruction method separates the two components well and suppresses the noise better. 
This is reflected in the average error of the reconstruction
\begin{equation}
    \eta_{M} = \sum_{i,j} \frac{\sqrt{(\Delta M_x^\prime(i,j))^2 +  (\Delta M_y^\prime(i,j))^2}}{N} 
\end{equation}
where $\Delta M_x^\prime = M_x - M_x^\prime$, $M_x$ is the true magnetization, $M_x^\prime$ is the reconstruction, $i$ and $j$ are the pixels indices and $N$ is the total number of pixels.
The single-component reconstruction has an average error of $\eta_{M} = 5.02\,\mu_B/$nm$^2$, compared to just $\eta_{M} = 1.06\,\mu_B/$nm$^2$, for the multi-component reconstruction.
This improvement can be quantified as the average reconstruction fidelity,
\begin{equation}
    F_M = 1 - \frac{\eta_M}{\overline{M}}
\end{equation}
where $\overline{M} = \sqrt{M_x^2 + M_y^2}/N$ is the average magnetization of the simulation and is only calculated within the masked region. 
The single-component reconstruction has a fidelity of $F = 0.57$ compared to $F = 0.91$ for the multi-component, demonstrating significant improvement. 
As such, we conclude that provided the decomposition of the vector magnetic components is possible without additional artifacts, it is always the preferred method for reconstruction. 

The multi-component reconstruction method works better because inhomogeneous magnetization distributions will produce x and y fields that don't match the smooth gradient that should exist above the material in these directions. 
As mentioned before the edge fields themselves have multiple solutions but these solutions won't necessarily match the fields away from the edges. 
The single-component contains stronger fields on the edge which results in these fields being reproduced first at the potential cost of the fields above the material. 
Having additional components effectively increases the error weighting of the xy-components, thus restricting the solution space.
It would be possible to produce a similar result by weighting the error functions with the mask mentioned previously. 
However, it is practically easier to perform the Fourier decomposition for each image than to dynamically produce masks with appropriately varying weights.

\section{Magnetic textures with 3D spin orientations}

\begin{figure*}
    \centering
    \includegraphics{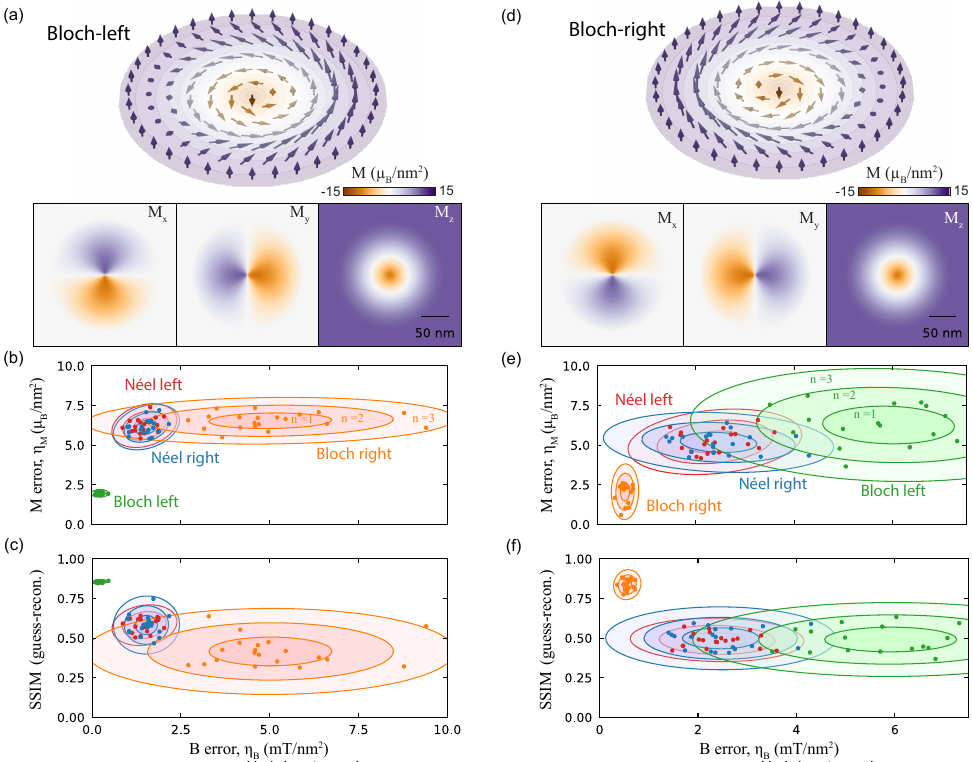}
    \caption{Identification of Bloch Skyrmion type 
    (a) Simulation of a left-handed Bloch Skyrmion.
    (b) Comparison of the magnetic field and magnetization error with different initial guesses of the Skyrmion type. Simulations were performed with magnetic field images with SNR = 10, with 20 simulations for 100 Epochs for each type. 
    The number $n = 1, 2, 3$ denotes the standard deviation region of each data set.
    (c) Comparison of the magnetic field error and the SSIM between the reconstructed in-plane magnetization and the expectation of the in-plane magnetization for the initial guess. 
    (d-f) Same as (a-c) but for a right-handed Bloch Skyrmion. 
    }
    \label{fig: bloch}
\end{figure*}

Magnetization directions are not always confined to a 2D plane and can in general point in all three directions. 
Determining the inner magnetic structure of such complex textures can be useful for understanding the physics of the material in which they exist. 
For example, topological spin textures like Skyrmions (Fig.~\ref{fig:skyrmion}) and Merons (Fig.~\ref{fig:meron}) can have an internal structure that varies in all directions.
Importantly this internal structure will change depending on the type (Bloch vs N\'eel) and the helicity of the quasi-particle.
The problem with the reconstruction of these textures is the 3D magnetic phase space is rich with local minima in the NN reconstruction and thus can return a result that is not the sought-after spin texture of the sample.

A potential application of the NN reconstruction would be the differentiation between topological spin textures. 
While Skyrmions and Merons yield significantly different stray-field profiles, the magnetic variation between Bloch, N\'eel Skyrmions (or domain walls), and their handedness is quite small.
This has been demonstrated for domain walls\,\cite{Tetienne2015} and for magnetic vortices\,\cite{Thiel2016}, where models are often fitted to the stray fields to differentiate between similar magnetic origins.
Unfortunately, the NN reconstruction by itself does not always converge to the correct spin-texture as there are many deep local minima in the solution space.

To overcome this limitation we introduce an initial magnetization distribution guess, which restricts the reconstruction to a region around the initial guess. 
In this way, the network output is modified such that
\begin{equation}
    M_{NN}^\prime = M_{NN} + \alpha M_{\rm init.} + \sigma
\end{equation}
where $ M_{\rm init.}$ is the initial guess for the magnetization in all three directions, $\alpha$ is a scaling factor, and $\sigma$ is random noise.
To minimise over-constraining the neural network by passing a perfect initial guess, the magnetization distribution is scaled to a fraction of the true value ($\alpha = 0.8$ in this work) and noise is added to each guess ($\sigma$ is set to 20\% noise here). 
This modification results in the NN output finding the difference between the initial guess and a better approximation of the magnetization. 
Consequently, the network will find a deep local minima near the initial guess, which can be used to identify if the initial guess was correct.

Initial guesses improve the reconstruction of topological spin textures and afford the ability to differentiate between the different types for an unknown spin texture. 
To illustrate this capability we simulated both left and right-handed Bloch skyrmions (Fig.~\ref{fig: bloch}\,a and d) and then compared the reconstruction given different initial guesses.
The Skyrmion is given a diameter of 100\,nm and the magnetic field was calculated at a standoff of 50\,nm with an SNR of $10$. 
The initial guess, $M_{\rm init}$, represents different magnetization structures that correspond to Skyrmions of different configurations: N\'eel left or right or Bloch left or right. 
For the example with a simulated Bloch left structure (a-c), we show how the error of the reconstruction varies when the NN receives different initial configurations $M_{\rm init}$. 
Fig.~\ref{fig: bloch}b shows errors in magnetic field and magnetization reconstruction for each simulation. 
The most accurate reconstruction occurs when both errors are minimal (lower left corner). 
This case occurs when the initial guess coincides with the underlying magnetization structure (Bloch-left) and is also true for Bloch-right type skyrmions (Fig.~\ref{fig: bloch}e).
This indicates that supplying the correct initial guess produces the best approximation of the magnetization and thus this can be used to determine the skyrmion type. 
Some examples of the reconstructed magnetization with different initial guesses are shown in appendix  Fig.~\ref{sifig: bloch recon}.

Importantly, the difference between the reconstructed and true magnetization used in Fig.~\ref{fig: bloch}b and e, is in general unknown and thus cannot be used for experiments. 
An alternative comparison metric that is available to experiments to compare the image reconstruction quality is the Structural Similarity (SSIM) Index\,\cite{zhou_ssmi_2004}, which more accurately represents the perceived similarity of the reconstruction by comparing the luminance, contrast, and structure of the reconstruction to the ground truth. 
The metric can also be used to determine the distance from the initial guess that the network had to travel in the reconstruction, where a good initial guess requires less optimisation to converge.
The combination of the SSIM between the reconstructed magnetization and the initial guess without modification ($M_{\rm init.}$) and the magnetic field error can produce an isolated reconstruction cluster with greater than 3$\sigma$ separation from other guesses (Fig.~\ref{fig: bloch} c and f).
This allows for the differentiation of both left and right-handed Bloch Skyrmions, even in the presence of noise.
In contrast, the same procedure for N\'eel skyrmions can determine the difference between N\'eel and Bloch type but fails to differentiate handedness (see appendix Fig.~\ref{sifig: neel recon}).

\section{Conclusion}

In this work, we have shown that NN magnetization reconstruction can be extended from a singular uniaxial direction in an image to performing magnetic reconstructions on images with non-uniform magnetization directions. 
We have shown that NN magnetization reconstruction can be improved significantly by implementing bounds like weighted masks and by further restricting the solution space by using all three vector components of the magnetic field. 
Both of these improvements cannot be implemented with the traditional Fourier space transformation approach, demonstrating the distinct advantage physically informed neural networks have by controlling the physical model to match the reality of the sample. 
Finally, we have demonstrated how these NNs can reconstruct non-trivial magnetization distributions in topological spin-textures and can be used to determine the type and handiness of Bloch Skyrmions.

\begin{acknowledgments}
This work was supported by the Australian Research Council (ARC) through grant DE230100192, and by the ERC consolidator grant project QS2DM. 

\end{acknowledgments}

\appendix

\section{Limiting over-fitting of noise}\label{App: overfitting}
It is generally a problem with a model like this that over-fitting will result in the 
noise of the image being fitted exactly by the variation in the reconstructed source in the process known as over-fitting.
While one can reduce the number of Epochs to minimise this, additional features can be included in the model that actively limit over-fitting.  

\textit{Standard deviation error function:} In addition to the normal error function, one can include a standard deviation error such that,
\begin{equation}
    \eta = \zeta_L\sum_i \left| B_i - B_i^\prime \right| + \alpha \text{ std}(M^\prime) 
\end{equation}
where $B$ is the measured field, $B^\prime$ and $M^\prime$ are the reconstructed magnetic field and magnetization respectively, and $\alpha$ is a scaling parameter that controls the extent to which the standard deviation should be minimized to decrease the error.

\textit{Spatial resolution modelling:} The spatial resolution of the sensor can be directly incorporated into the model by introducing a Gaussian or Lorentzian convolution with the reconstructed magnetization. 
This inclusion is particularly relevant for fully-connected neural networks which contains no spatial averaging in the network itself.
The output of the network is modified such that
\begin{equation}
    M = f_\sigma \ast M'
\end{equation}
where $f_\sigma$ is the spatial filter that is relevant to the measurement.
In the case of scanning systems, the spatial resolution is often defined by the standoff, which has a Lorentzian spatial filter with a width $\sigma = z^{\prime} /2$. 
While for wide-field system the spatial resolution is commonly limited by the diffraction limit, which has a Gaussian spatial filter with a width $\sigma = \lambda /2$NA, where $\lambda$ is the relevant wavelength and NA is the numerical aperture of the objective.

\section{Non-aligned magnetic and image axes}
The x-y axes of the measurement image and the direction of the in-plane magnetization are often not aligned, i.e. the magnetization lies somewhere between the x- and y-axis.
In these conditions, rotating the $B_{xyz}$ into the sample's frame $B_{x'y'z}$ and using these magnetic field components for the reconstruction can lead to further improvement due to the higher spatial correlations between $B_{x'y'}$ and $M_{x'y'}$ in the sample frame\,\cite{healey_varied_2022} compared to the image frame. 
The transformation back into the magnetic field is still performed in the image frame, which requires rotating the output magnetization into the image frame, transforming it into the magnetic fields, and then rotating these fields into the sample frame for comparison.

\section{Reliable estimation of magnetization}
For out-of-plane magnetization, an appropriate sample design to minimise edge magnetic material and maximise SNR leads to a good estimate of the magnetization without the need of masks or NNs.
However, in the case of in-plane magnetization, weighted masks become more necessary.
Additionally, a reliable estimation of the magnetization of a sample is nominally only possible with scanning probe measurements, as the point spread function of wide-field approaches generally leads to an underestimation of the magnetic field and thus an underestimation of the magnetization~\cite{scholten_aberration_2022, nishimura_investigations_2024}. 
While it may be possible to engineer the point spread function of such widefield approaches, it has yet to be demonstrated for improved magnetic field estimation, rather it has only been investigated in terms of improved sensitivity~\cite{McCloskey2020}.

\section{Model and network architecture}\label{App: CNN arch}
The convolutional neural network used throughout this work follows a U-net architecture where and additional branch is added from the central node to account for the multiple outputs. 
This approach was found to produce a more reliable reconstruction than separating the outputs at the end of the U-net. 
In particular, the latter approach often leads to correlated erroneous magnetization reconstructions because the degrees of freedom between the two outputs are too small to completely decouple the results. 
The layers architecture for the CNN is displayed in Table \ref{table: CNN architecture}.

	\begin{table}
		\centering
		\begin{tabular}{c  c  c  c c}
            \hline
            \hline
			& \, Layer Type \,  & \, Filters \, & \, Kernel Size \, & \, Stride \, \, \\
			\hline
			& conv & 8 & 5 & 1 \\
			& conv & 8 & 5 & 2 \\
			& conv & 16 & 5 & 2  \\
			& conv & 32 & 5 & 2 \\
			& conv & 64 & 5 & 2  \\
			& conv & 128 & 5 & 2\\
            \hline 
			& conv & 64 & 5 & 2\\
			& conv & 32 & 5 & 2\\
			output & conv  & 16  & 5 & 2 \\
			1 & conv & 8 & 5 & 2\\
			& conv & 1 & 5 & 1 \\
			& conv & 1 & 5 & 1 \\
			\hline
            & conv & 64 & 5 & 2\\
			& conv & 32 & 5 & 2\\
			output & conv  & 16  & 5 & 2 \\
			2 & conv & 8 & 5 & 2\\
			& conv & 1 & 5 & 1 \\
			& conv & 1 & 5 & 1 \\
			\hline
		\end{tabular}
		\caption{Architecture of the convolutional neural network}
		\label{table: CNN architecture}
	\end{table}
	\vspace{0.8\baselineskip}

\subsection{Bayesian regularisation method}
An alternative approach to the Fourier-based reconstruction method without neural networks is Bayesian estimation.
In these approaches, the ill-poised transformation is approximated using a regularization parameter that eliminates the $1/0$-terms due to its non-zero addition to the matrix before inversion and can be used to model the noise in the image.   
As such, this approach combines data fitting and noise rejection through regularisation~\cite{Meltzer2017, Clement2019, midha_optimized_2024}.
While currently these approaches have only been demonstrated for current density, they are trivially extended to magnetization and show a similar performance improvement. 
However, the resulting source image strongly depends on the regularisation parameter which needs to be tuned to limit over-smoothing while maintaining good noise reduction~\cite{midha_optimized_2024}. 
This process can be complex and may not apply to all datasets where the resulting source distribution is unknown or it is difficult to eliminate certain outcomes as invalid.

\section{Reconstruction of non-trivial in-plane magnetization}

\begin{figure}
    \centering
    \includegraphics{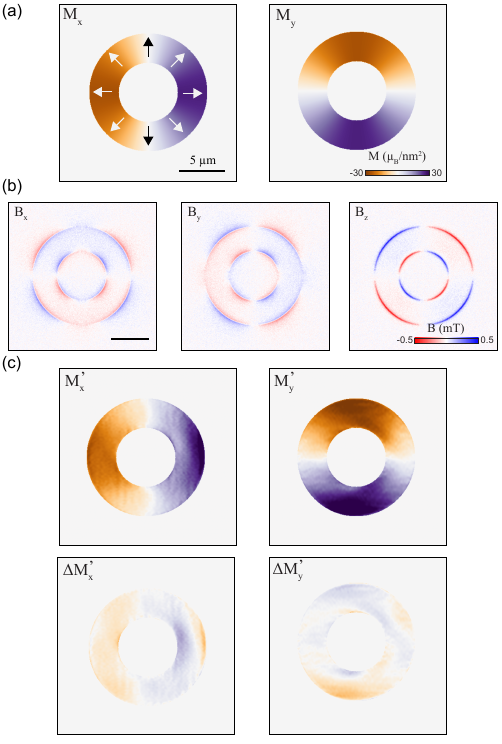}
    \caption{Reconstruction of rotating magnetization direction.
    (a) Simulation of a torus magnetic sample with a magnetization direction normal to the surface.
    (b) Simulated magnetic field from (a) with an SNR of 10. 
    (c) Magnetic reconstructed from the fields in (b) (top) and the difference between the reconstructed and true value (bottom).
    }
    \label{fig:torus}
\end{figure}

We show how NN reconstruction can be used to reconstruct non-trivial magnetization distributions. 
To demonstrate this, we simulate a torus with a rotating magnetization direction, such that the magnetization direction is always pointing radially outwards and has a magnitude of $M = 25 \mu_B/$nm$^2$ (Fig.~\ref{fig:torus}a). 
We use the multi-component reconstruction approach with randomised noise (SNR = 10, Fig.~\ref{fig:torus}b).
Even in the presence of noise, the NN can reconstruct this non-uniform magnetization distribution (Fig.~\ref{fig:torus}c).

The difference in reconstructed magnetization and the ground truth gives an average difference of $\eta_m = 5~\mu_B/$nm$^2$ and a fidelity of $F = 0.8$, dominated by a gradient in the magnetization. 
Additionally, we find that the neural network approach can achieve a high level of similarity with values of SSIM($M_x$) = 0.986 and SSIM($M_x$) = 0.969.
While both these metrics indicate that the reconstruction is relatively good, the interpretation of such reconstructions needs to be taken with some caution. 
For instance, one may incorrectly conclude that there is a gradient in the magnetization strength from this reconstruction, indicating a stronger magnetization stability at the outer edge compared to the inner edge.

\section{Skyrmion reconstruction}

\begin{figure}
    \centering
    \includegraphics{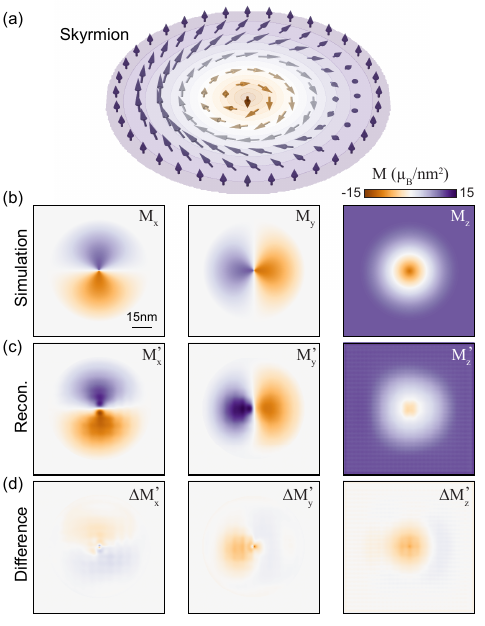}
    \caption{Reconstruction of a topological spin texture -- the Skyrmion. 
    (a) Depiction of the spin direction inside of a skyrmion.
    (b) Simulation of the magnetization for a  Bloch Skyrmion with a spin-up background, where the Skyrmion has a diameter of 100\,nm and was simulated with a standoff of 10\,nm.
    (c) Reconstruction of the magnetization from the simulation in (b).
    (d) The difference between the reconstruction (c) and the ground truth (b).}
    \label{fig:skyrmion}
\end{figure}

To illustrate the reconstruction issue without initial guesses, we simulate a Bloch-type Skyrmion\,\cite{zhang_skyrmion-electronics_2020} embedded in a background of homogenous spin-up magnetization (Fig.~\ref{fig:skyrmion}b). 
The Skyrmion is given a diameter of 100\,nm and the magnetic field was calculated at a standoff of 10\,nm. 
To account for the three-dimensional magnetization direction we introduce a third output to the NN so that all cartesian components of the magnetization are accounted for. 
The simulated magnetization (Fig.~\ref{fig:skyrmion}b) and the reconstructed (Fig.~\ref{fig:skyrmion}c) share the same overall structure which is evident in the difference between the two (Fig.~\ref{fig:skyrmion}d). 
This indicates that under the right conditions, untrained NNs can be used to reconstruct 3D varying magnetic textures, however, they are still prone to falling into local minima and may therefore not be effective in analysing all spin textures.

\section{Meron reconstruction}
\begin{figure}
    \centering
    \includegraphics{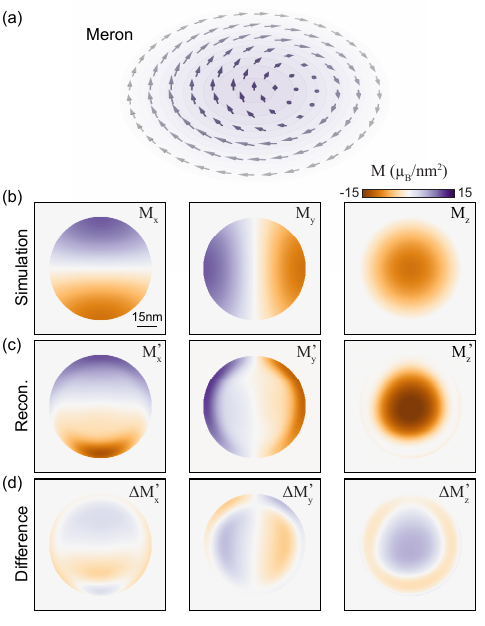}
    \caption{Reconstruction of a topological spin texture -- Meron. 
    (a) Depiction of the spin direction inside of a Meron.
    (b) Simulation of the magnetization for a spin-up Meron with no net magnetization background, i.e., a pure XY-model system with no long-range order. The Meron has a diameter of 100\,nm and the magnetic field was simulated with a standoff of 10\,nm.
    (c) Reconstruction of the magnetization from the simulation in (b).
    (d) The difference between the reconstruction (c) and the ground truth (b).}
    \label{fig:meron}
\end{figure}

We also simulated a spin-up Meron embedded in a background of zero magnetization, akin to a Meron in an antiferromagnetic XY-model magnet with no long-range order (Fig.~\ref{fig:meron}b). 
The simulated Meron magnetization (Fig.~\ref{fig:meron}b) and the reconstructed magnetization (Fig.~\ref{fig:meron}c) share a general shape that is consistent but fails to capture the full magnetization distribution.
This is evident in the difference (Fig.~\ref{fig:meron}d), which highlights that the reconstruction has preferentially allocated magnetization to the edge of the Meron similar to the Torus reconstruction. 
While this reconstruction could be used to differentiate between different types and orientations of Merons, it is not reliable enough to probe the internal magnetization rotation itself. 

\section{Additional reconstruction comparison}
\begin{figure*}
    \centering
    \includegraphics[width=\linewidth]{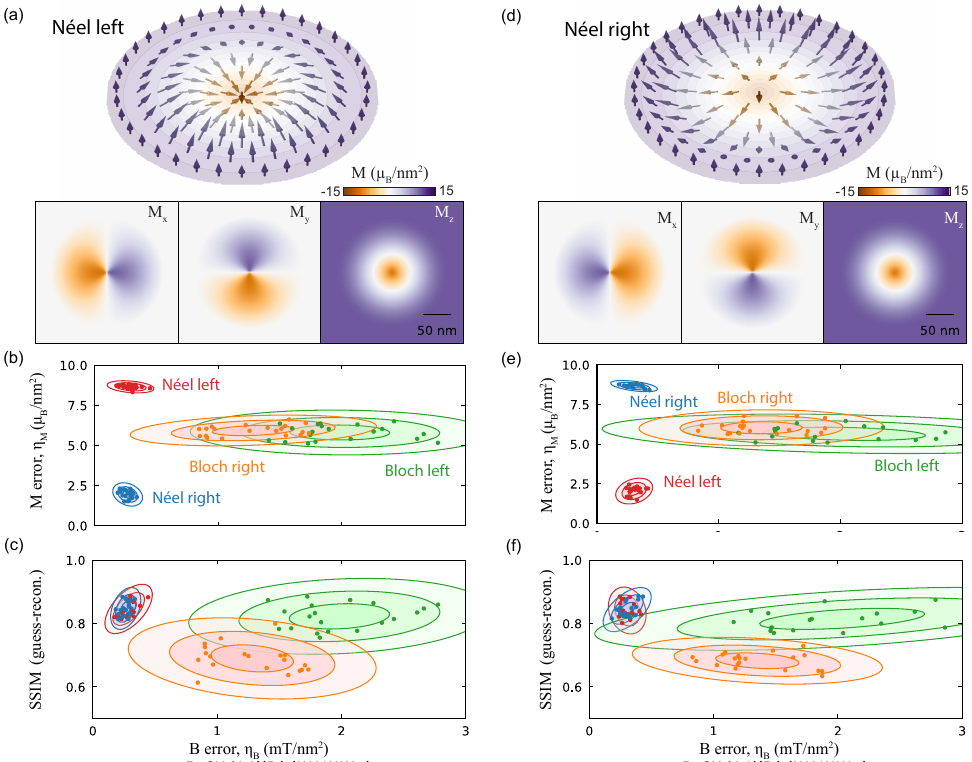}
    \caption{Example reconstructions type determination for N\'eel type skyrmions. 
    (a) Simulation of a left-handed N\'eel Skyrmion.
    (b) Comparison of the magnetic field and magnetization error with different initial guesses of the Skyrmion type. Simulations were performed with magnetic field images with SNR = 10, with 20 simulations for 100 Epochs for each type. 
    (c) Comparison of the magnetic field error and the SSIM between the reconstructed in-plane magnetization and the expectation of the in-plane magnetization for the initial guess. 
    (d-f) Same as (a-c) but for a right-handed N\'eel Skyrmion. 
    }
    \label{sifig: neel recon}
\end{figure*}

We have also performed reconstructions of N\'eel skyrmions in the same fashions as the Bloch skyrmions (Fig.~\ref{sifig: neel recon}). 
Unlike the Bloch skyrmions, the difference between the handedness for N\'eel type is not easily distinguished.

\begin{figure}
    \centering
    \includegraphics[width=\linewidth]{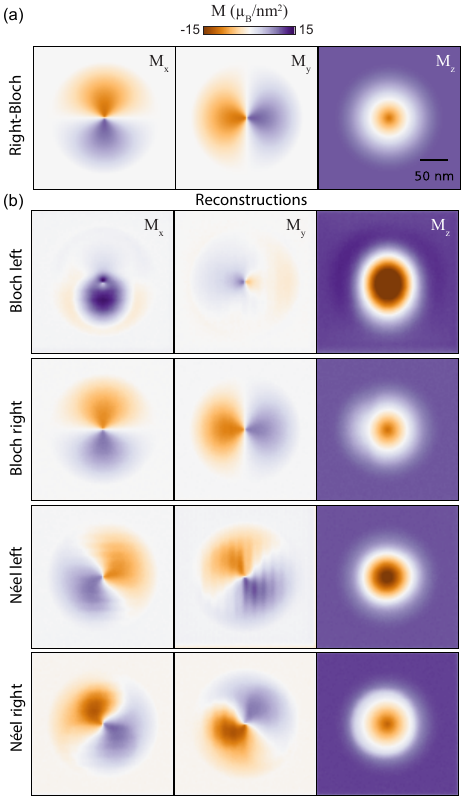}
    \caption{Example reconstructions with different initial guesses for a right-handed Bloch skyrmion. 
    (a) The magnetization ground truth. 
    (b) magnetization reconstructions from different initial guesses.  }
    \label{sifig: bloch recon}
\end{figure}

\bibliography{bib.bib}

\end{document}